\documentclass[aps,floatfix,superscriptaddress,notitlepage,nofootinbib]{revtex4-1}
\usepackage{amsmath,amssymb,amsfonts,graphics,graphicx,dcolumn,bm,enumerate}
\usepackage{comment,natbib,appendix}
\usepackage{multirow,color}
\usepackage{chngpage}
\usepackage{afterpage}
\usepackage{xcolor}
\usepackage{amsthm}
\usepackage{natbib}
\usepackage{hyperref}
\usepackage[margin=1.2in]{geometry}
\usepackage{epstopdf}
\usepackage{float}
\usepackage{amsmath}

\newcommand{\cmp}
{\affiliation{Saha Institute of Nuclear Physics, Kolkata 700064, India.}}
\newcommand{\isi}
{\affiliation{Economic Research Unit, Indian Statistical Institute, Kolkata 700108, India.}}
\newcommand{\raghunathpur}
{\affiliation{Department of Physics, Raghunathpur College, Raghunathpur, Purulia 723133, India.}}

\begin{document}
\title{Scaling and Kinetic Exchange Like Behavior of Hirsch
Index and Total Citation Distributions: Scopus-CiteScore
Data Analysis \footnote{Dedicated to the loving memory of Prof. Amit Dutta, IIT Kanpur, India.}}
\author{Asim Ghosh}
\email[Email: ]{asimghosh066@gmail.com}
\raghunathpur
 
 \author{Bikas K. Chakrabarti }%
 \email[Email: ]{bikask.chakrabarti@saha.ac.in}
 \cmp \isi 
 
\begin{abstract}
We analyze the data distributions $f(h)$, $f(N_c$) and $f(N_p)$ of the
Hirsch index $(h)$, total citations ($N_c$) and total number of papers
($N_p$) of the top scoring  120,000 authors (scientists) from the
Stanford cite-score (or c-score) 2022 list and their corresponding  $h$ ($3 \le h
\le 284$), $N_c (1009 \le N_c \le 428620$) and $N_p$ ($3\le N_p \le 3791$)
statistics from the Scopus data. For reasons
explained in the text, we divided the data of these top scorers (c-scores
in the range 5.6125 to 3.3461) into six successive equal-sized Groups of
20,000 authors or scientists. We tried to fit, in each Group, $f(h)$, $f(Nc)$
and $f(Np)$ with Gamma distributions, viewing them as the   ``wealth
distributions" in the fixed saving-propensity kinetic exchange models and
found $f(h) \sim h^{\gamma_h} \mathrm{exp} (-h/T_h)$  with fitting noise level or
temperature level ($T_h$) and average value of $h$, and the power $\gamma_h$
determined by the ``citation saving propensity” in each Group. We further showed that using some earlier proposed power law scaling like  $h = D_c N_c^{\alpha_c}$ (or $h = D_p N_p^{\alpha_p}$) with $\alpha_c = 1/2 = \alpha_p$, we can derive the observed $f(h)$ from the observed $f(N_c)$ or
$f(N_p)$, with $D_c = 0.5$, but $D_p$ depending on the Group considered. This observation suggests that the average citations per paper ($N_c/N_p$) in each group ($= (D_p/D_c)^2 =4D_p^2$)  vary (from 58 to 29) with the c-score range of the
six Groups considered here, implying different effective Dunbar-like
coordination numbers of the scientists belonging to different groups
or networks.
\end{abstract}

\maketitle

\section{Introduction}
A popular measure of the success of individual scientist or author (called
scientist here generally) has been the  Hirsch Index \cite{1Hirsch2005} or h-index, which
can be viewed as the fixed point \cite{2Ghosh2022}  of the non-linear function relating
the monotonically decreasing number of publications  ($n_p$) with
increasing number of citations ($n_c$): $n_p = h = n_c$ of the scientist.
Mapping the citation function to a  combinatorial Fermi one, Yong
proposed \cite{3Yong2014} the relationship
\begin{equation}
    h = D_c N_c ^ {\alpha_c},
    \label{eq:1}
\end{equation}
\noindent with $D_c \simeq 0.54$  and $\alpha_c = 1/2$ for any scientist with
Hirsch index value $h$ and total citations $N_c = \sum_p n_c$ from all his
or her publications (denoted by $p$), in the limit $N_c \rightarrow \infty$.
Several attempts to check the validity of such a relationship
between $h$ and $N_c$ have been made, see e.g., Redner \cite{4Redner2010} (supporting
the relation (\ref{eq:1}) with the exponent $\alpha_c$ value equal to 0.5, from the
data analysis for 255 scientists) and  Radicchi and Castellano \cite{5Radicchi2013} (
analysing a much larger set of data for 83,897 scientists) who  found
 the best fit value of the exponent  $\alpha_c  \simeq 0.42$. Ghosh et
al. \cite{2Ghosh2022} studied the Widom-Stauffer like scaling behavior of the Hirsch
index for the fiber bundle  as well as percolation models away from the
``critical'' breaking point or stress and percolation point respectively and
proposed
\begin{equation}
  h \sim  \sqrt N_c / [\mathrm{log} N_c], 
  \label{eq:2}
\end{equation}
for the citations of individual scientists, giving reasonable agreement
with the google scholar data for 1000 scientists (with  $h$-indices in the
range $17 \le h \le 221$  and total number of citations  $N_c$ in the
range $996 \le Nc \le 348680$).

\begin{figure}
    \centering
    \includegraphics[width=\textwidth]{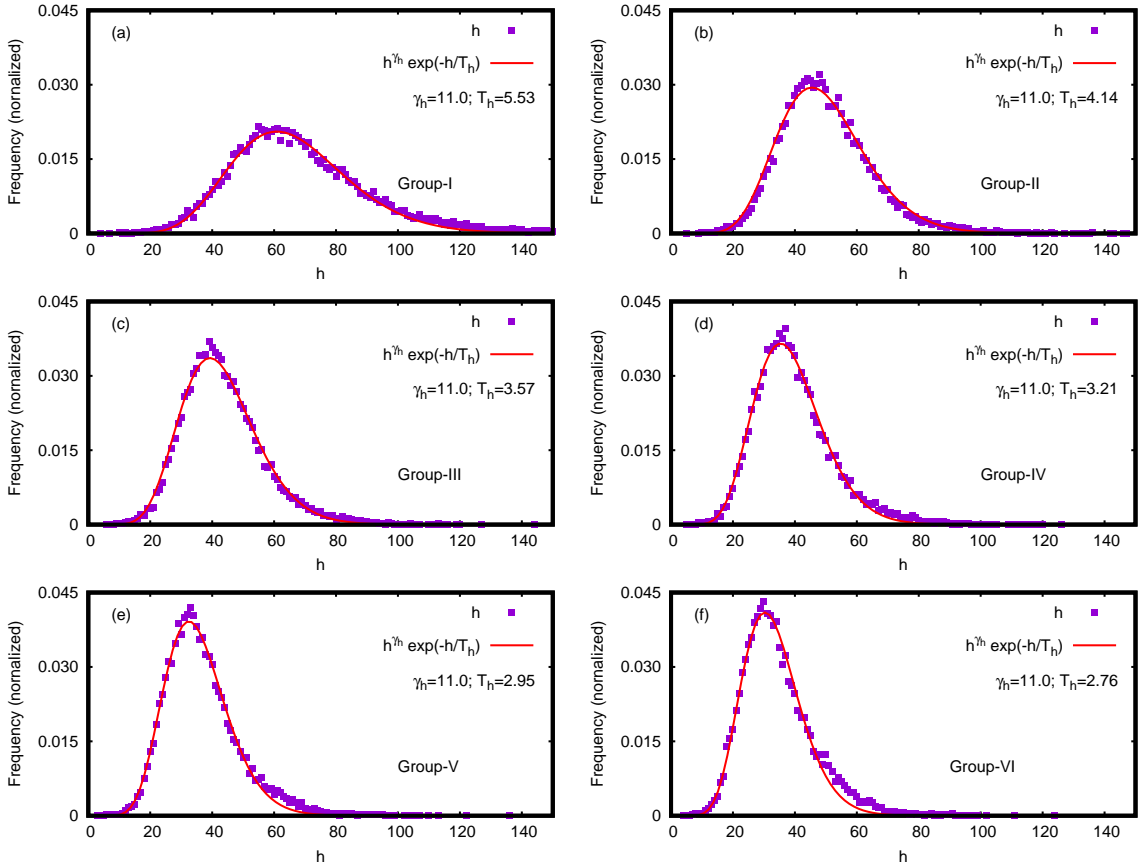}
    \caption{Frequency distribution (normalized) for the Hirsch index ($h$) of the authors in Groups I to VI, shown in Figs. 1a to 1f. The fitting Gamma functions are also shown.}
    \label{fig:1}
\end{figure}

\begin{figure}
    \centering
    \includegraphics[width=\textwidth]{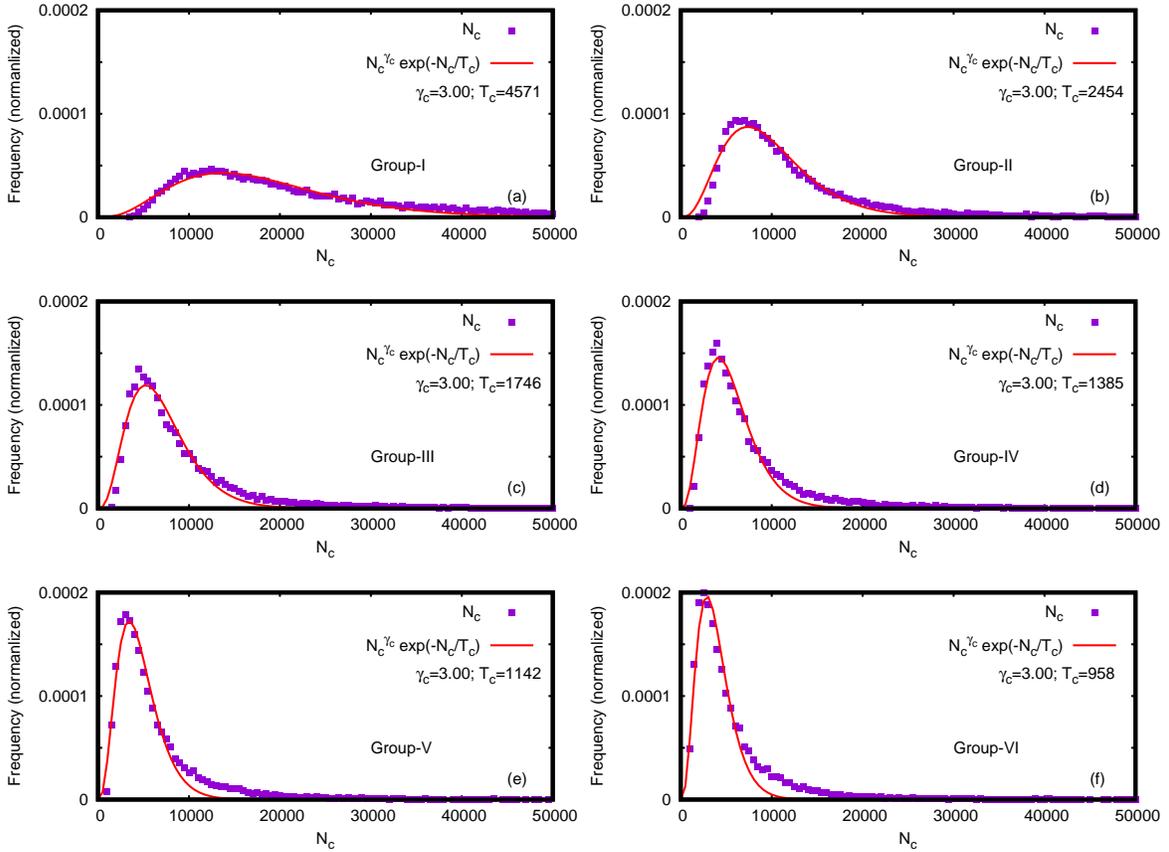}
    \caption{Frequency distribution (normalized) of the total number of citations ($N_c$)  of the authors in Groups I to VI, shown in Figs. 2a to 2f. The fitting Gamma functions are also shown.}
    \label{fig:2}
\end{figure}

\begin{figure}
    \centering
    \includegraphics[width=\textwidth]{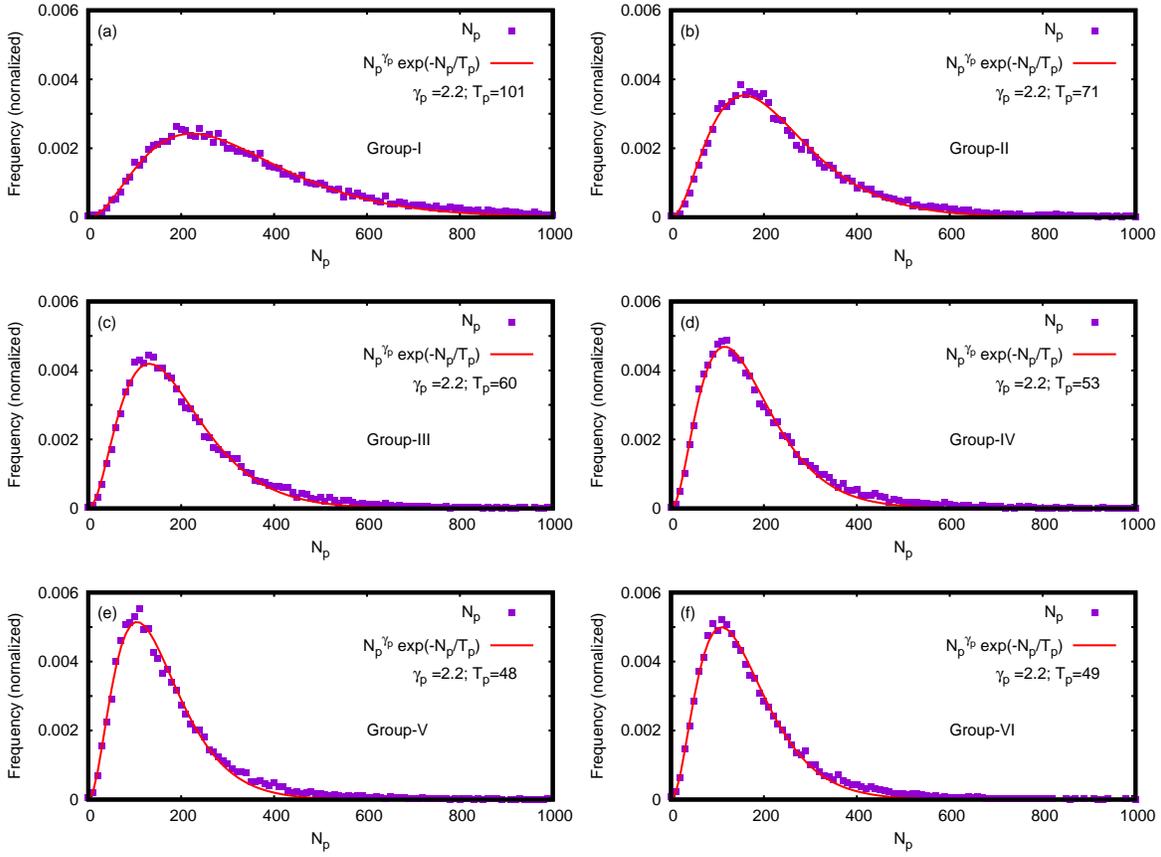}
    \caption{Frequency distribution (normalized) of the total number of papers ($N_p$)  of the authors in Groups I to VI, shown in Figs. 3a to 3f. The fitting Gamma functions are also shown.}
    \label{fig:3}
\end{figure}

\begin{figure}
    \centering
    \includegraphics[width=1.0\textwidth]{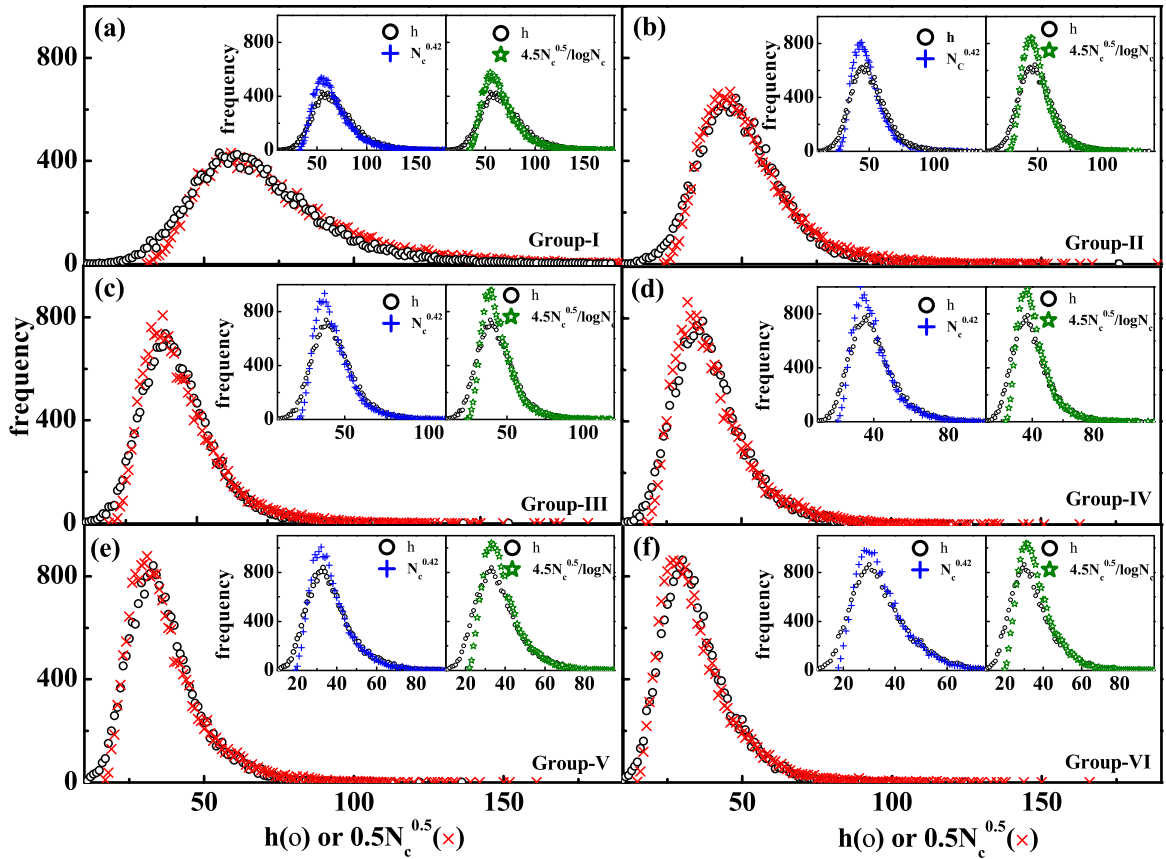}
    \caption{Frequency distribution for Hirsch
index ($h$), obtained directly from the data source
(denoted by symbol o) and that obtained from the
number of total citations ($N_c$) using relation (\ref{eq:1})
with $\alpha_c = 0.5$  (Yong \cite{3Yong2014}) and $D_c = 0.5$
(denoted by symbol $\times$). The observed overlaps
confirm the relation (\ref{eq:1}) with $\alpha_c = 0.5$. The
insets show the same with $\alpha_c = 0.42$ \cite{5Radicchi2013} (left)
and that with $\alpha_c = 0.5$ with an inverse $\mathrm{log} N_c$
correction \cite{2Ghosh2022} (right); there seem to be
considerable  mismatches.}
    \label{fig:4}
\end{figure}

\begin{figure}
    \centering
    \includegraphics[width=1.\textwidth]{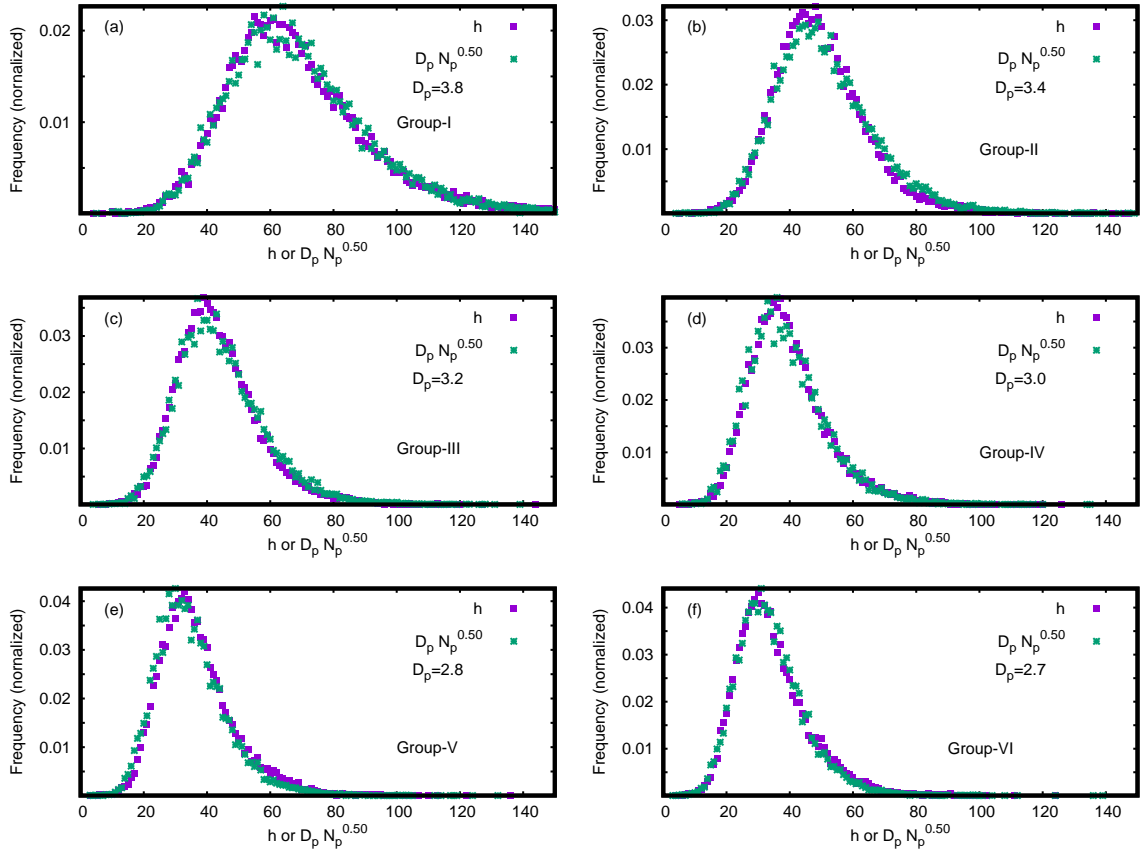}
    \caption{Frequency distribution of Hirsch index ($h$) obtained directly from the data source and those obtained from the number ($N_p$) of total publication by the authors of the Group I to VI using relation (\ref{eq:5}) $\alpha_p=0.5$}
    \label{fig:5}
\end{figure}

\begin{figure}
    \centering
    \includegraphics[width=0.8\textwidth]{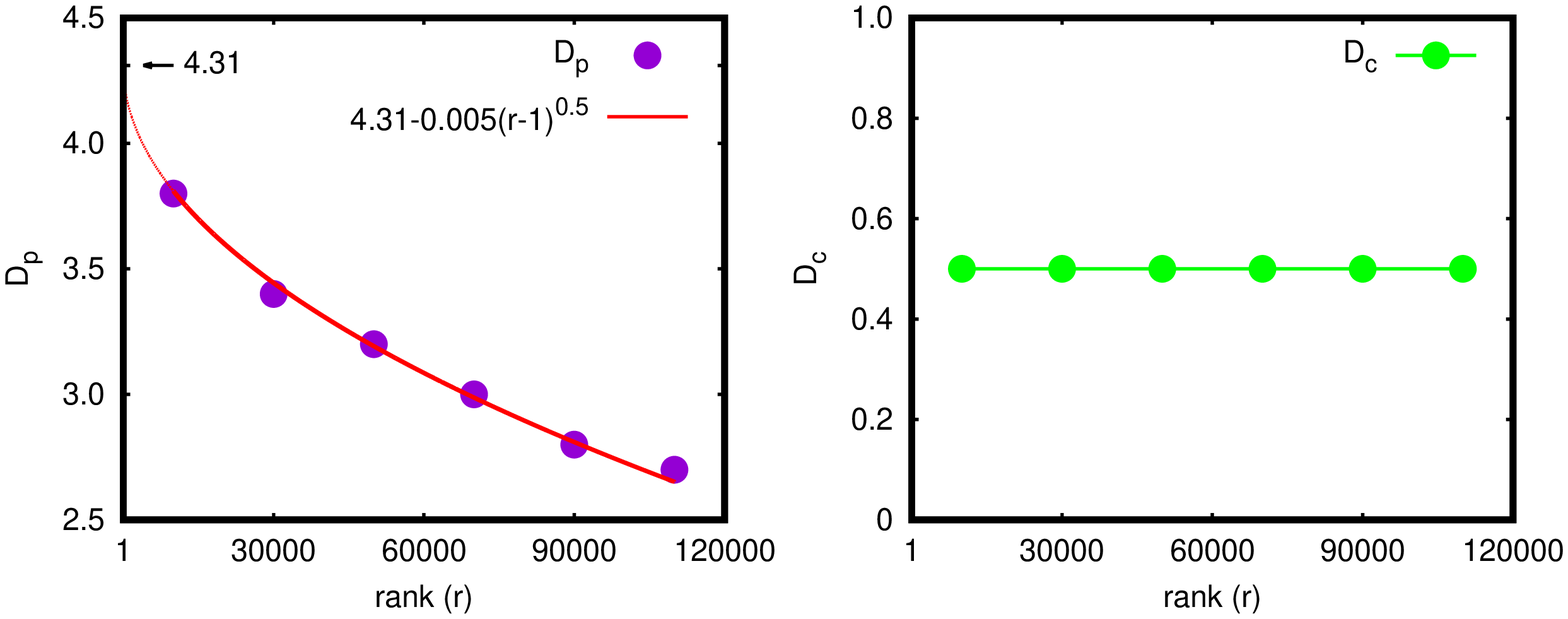}
    \caption{Extrapolated (statistical) value of  the prefactor $D_p$ in eqn. (\ref{eq:5})
for top ranking ($r = 1$) scientist. We plot the values of $D_p$ for
different Groups  from table  \ref{tab:2} against the middle rank  ($r$) of the
corresponding rank range (the same plot for the prefactor $D_c$ in Eqn. (\ref{eq:1}) remains a horizontal line) .}
    \label{fig:6}
\end{figure}

We find here, in each of the six equal-size Groups of twenty thousand top ranking scientists from the Elsevier Stanford c-score list \cite{Ioannidis2019,6scopus,7elsevier} (total one hundred and twenty thousand top cited scientists), the distributions (frequencies) $f(h)$, $f(Nc)$ and $f(Np)$ of their Hirsch index ($h$), total citations ($N_c=\sum_p n_c$) and total number of papers ($N_p=\sum_p n_p$) all fit very well with Gamma function form:

\begin{subequations}\label{eq:3}
\begin{equation}\label{eq:3a}
    f(h) \sim h^{\gamma_h} [\mathrm{exp}(-h/T_h)],
\end{equation}
\begin{equation}\label{eq:3b}
  f(N_c) \sim N_c^{\gamma_c} [\mathrm{exp}(-N_c/T_c)],  
\end{equation}
\begin{equation}\label{eq:3c}
  f(N_p) \sim h^{\gamma_p} [\mathrm{exp}(-N_p/T_p)],  
\end{equation}
\end{subequations}
\noindent with the exponent values  $\gamma_h \simeq 11.0$, $\gamma_c \simeq
3.0$,  $\gamma_p
\simeq 2.2$, and the noise levels $T_h, T_c, T_P$, dependent on the
c-score range generally decreases with decreasing  c-score (see Figs. \ref{fig:1}, \ref{fig:2} and \ref{fig:3} in the next section on data analysis). We tried to
obtain $f(h)$ from $f(N_c)$ using the power law
relation (\ref{eq:1}) and got excellent fit (for $D_c =
0.5$ and $\alpha_c = 1/2$; see Figs. \ref{fig:4}). We also
tried obtain $f(h)$ from $f(N_p)$, using the relation
$h = D_p N_p^{\alpha_p}$ \cite{2Ghosh2022}, with $\alpha_p = 1/2$
but $D_p$ value depending on the Group. Again we
got good fit (see Figs \ref{fig:5}).

As mentioned above already, the values of
$h$ calculated from the corresponding values
of $N_c$, using relation (\ref{eq:1}) proposed by Yong
\cite{3Yong2014}, but with $D_c = 0.5$ (not with $D_c \simeq
0.54$ as suggested by Yong), gives
extremely good fit to the distributions $f(h)$
as shown in Figs. \ref{fig:1}. Other scaling relations $h
\sim N_c^{0.42}$ suggested by Radicchi and Castellano \cite{5Radicchi2013} in 2013, or $h \sim N_c^{0.50}/ \mathrm{log} N_c$
suggested by Ghosh et al. \cite{2Ghosh2022} in 2022 do not give comparable good fits.

In addition, we find the $T_h$ values for each Group (the six c-score
ranges I-VI),
calculated using  the relation
\begin{equation}\label{eq:4}
  T_h = h_{av}/(\gamma_h + 1),  
\end{equation}
\noindent where $h_{av}$ denotes  the average of $f(h)$ in each Group, compares
very well with
the observed values. This relation suggests a
strong  correlation of the Chakraborti-Chakrabarti model \cite{8Chakraborti}  of ``wealth''
distribution where a fixed saving fraction of the wealth (which determines the
exponent value of $\gamma$ in the resulting Gamma distribution of wealth) is
retained in each kinetic exchange or interaction (see \cite{9Sen2014,10Patriarca2004}). If we consider  a similar
stochastic dynamics of paper citations, where the fixed fraction of
(confident or
core Group) ``citations'' (like wealth) in each paper-writing
(interaction) determines
the exponent $\gamma_h$  value and the corresponding noise level $T_h$  in the
resulting Gamma distribution $f(h)$ of the (wealth) h-index. The
equivalent wealth
conservation may be assumed to come (see e.g., \cite{Pareschi2014}) from the overall
constancy (node
coordination  number) of the citation network, discussed later.
In other words,  as in the  model \cite{8Chakraborti} of
``wealth” distribution, discussed above,  where a fixed saving fraction of the wealth
(which determines the
value of the exponent $\gamma_c$ in the Gamma distribution) is retained in
each kinetic exchange
or interaction, suggests here a similar stochastic dynamics of paper
citations, where the fixed
fraction of (confident or core Group) ``citations” (or wealth) in each
paper-writing (interaction)
determines the exponent $\gamma_c$ value and the corresponding noise level $T_h$
in the Hirsch index
$f(h)$ distributions in each Group.

We also found  an interesting feature of the citation network. As we
mentioned in connection with relation (\ref{eq:3c}), using the scaling relation 
\begin{equation}\label{eq:5}
  h=D_p N_p^{\alpha_p},  
\end{equation}
\noindent with  $\alpha_p = 0.5$, we get $D_p \simeq$ 3.8, 3.4, 3.2, 3.0, 2.8 and 2.7 respectively for the successively decreasing six c-score Groups I to VI.  Comparison of the relations
(\ref{eq:3a}) and (\ref{eq:3c}) with $\alpha_c = 0.5 = \alpha_p$, and as discussed above the best fit value of $D_c=0.5$, suggests the value of the  average
citation per paper 
$N_c/N_p$ for any of these scientists depends on average
on the Group in which the scientist belongs, and is  given
by $(D_p/D_c)^2 = 4D_p^2$,  ranging from 58 (for Group I) to 29
(for Group VI). This may be identified as the effective social
coordination number (or  Dunbar number \cite{Ghosh2021,Dunbar1992})  giving
the group size of the scientists, as in the primates.

\section{SCOPUS DATA ANALYSIS}
As mentioned already, we analyzed here the Elsevier Scopus \cite{6scopus} data for the Hirsch index $h$ and
the corresponding number $N_c$  of total citations for 120,000 scientists
who came at the top of Stanford c-score list \cite{7elsevier} last year (2022). We
divided the set into six equal Groups of 20,000 scientists having c-sore
rank ranges I [1-20000], II [20001-40000], III [40001-60000], IV
[60001-80000], V [80001-100000], and VI [100001-120000]. We observed that
for the scientists in each of these ranges, both the $h$-index values and
the total citations numbers $N_c$ have similar Gamma-like distributions
(see Fig. \ref{fig:1} (\ref{fig:1}a to \ref{fig:1}f) for distributions of $h$ and of $N_c$  for the six
ranges of c-score ranks mentioned above). We observe (see Figs \ref{fig:2}; \ref{fig:2}a to \ref{fig:2}f)
that the $h$-index distribution $f(h)$ in each of these six score ranges
fit very well to the Maxwell-Boltzmann like Gamma function form 3(a) with  $\gamma_h \simeq 11$ and the effective noise (temperature)
decreases  with increasing range (from I through VI). In Table \ref{tab:1}, we give
for each of the six ranges (I-VI) the estimated values of the most probable
value of Hirsch index $h^{mp}$, its average $h_{av}$ and the respective
noise level or temperature $T_h$.

\begin{table}[h]
    \centering
    \caption{Hirsch index ($h$) data fitting parameters obtained from relation (\ref{eq:3a}). }
    \begin{tabular}{|c|c|c|c|c|c|c|}
    \hline 
    Group & c-score rank & $\gamma_h$ & $h^{mp}$ & $h_{av}$ & $T_h$ & $h_{av}/ (\gamma_h + 1$)\\
    \hline 
 I & 1-20K & 11.0 & 61.0 & 68.8 & 5.53 &  5.73 \\
 \hline
II & 20k-40K  & 11.0  & 45.5  & 49.8 & 4.14 & 4.15\\
\hline
 III & 40K-60K  & 11.0 & 39.0  & 43.4 & 3.57 & 3.61\\
 \hline 
 IV & 60K-80K  & 11.0 &   35.0 & 39.5 & 3.21 & 3.29\\
 \hline 
 V & 80K-100K    & 11.0 & 32.5 & 36.6 & 2.95 & 3.05\\
 \hline 
VI & 100K-120K   & 11.0  & 29.9 & 34.5 & 2.76 & 2.88\\
\hline 
    \end{tabular}
    \label{tab:1}
\end{table}

\begin{table}[h]
    \centering
    \caption{Fitting parameters for the total number of citations ($N_c$) and the total number of papers ($N_p$) obtained using the relations (\ref{eq:1}) and (\ref{eq:5}).}
    \begin{tabular}{|c|c|c|c|c|c|c|}
    \hline 
    Group & c-score rank & $\alpha_c$ & $\alpha_p$ & $D_c$ (from relation (\ref{eq:1})) & $ D_p$ (from relation (\ref{eq:5})) & $ (D_p/D_c)^2$\\
    \hline 
I & 1-20K & 0.5 & 0.5 & 0.5 & 3.8 &  58 \\
 \hline
II & 20k-40K & 0.5 & 0.5 & 0.5 & 3.4 &  46 \\
\hline
 III & 40K-60K  & 0.5 & 0.5 & 0.5 & 3.2 &  41 \\
 \hline 
 IV & 60K-80K  & 0.5 & 0.5 & 0.5 & 3.0 &  36 \\
 \hline 
 V & 80K-100K    & 0.5 & 0.5 & 0.5 & 2.8 & 31  \\
 \hline 
VI & 100K-120K   & 0.5 & 0.5 & 0.5  & 2.7 & 29  \\
\hline 
\hline 
    \end{tabular}
    \label{tab:2}
\end{table}

In Figs. \ref{fig:4} (\ref{fig:4}a to \ref{fig:4}f) we compare the above-mentioned observed distribution
of $f(h)$ with those obtained by using  the Yong's scaling relation (\ref{eq:1}) with $\alpha = 0.5$ and
the best fit value of  $D_c = 0.5$ for the six different ranges I to VI of c-score
ranks. The overlap seems to be very good and encouraging. In contrast, insets of
Figs. \ref{fig:4}  we compared the same  $h$-index distributions $f(h)$ in
the six different ranges of c-score ranks with the $h$ values obtained the
$N_c$ values using relation (\ref{eq:1}) with $\alpha = 0.42$ (as observed in \cite{5Radicchi2013},
mentioned above) and the best fit value  of the prefactor. The
level of misfit is obvious. The same is true when one uses the relation
(\ref{eq:2}) between $h$ and $N_C$  with the best fit value ($4.5$) of the
prefactor (as suggested in \cite{2Ghosh2022}). Again the distributions
of $h$ and those obtained using relation (\ref{eq:2}) do not match (see the insets of Fig. \ref{fig:4}).

Our analysis (see Figs. \ref{fig:2})  for the Hirsch indices and the corresponding
values of the total citations (from Scopus data) for the top ranking
c-score authors therefore confirms the relation (\ref{eq:1}) with the exponent
$\alpha_c = 1/2$, as obtained by Yong \cite{3Yong2014}. This is because of the lack of matches (inset of the Figs. \ref{fig:4}) with $\alpha_c=0.42$ \cite{5Radicchi2013}  or $\alpha_c=1/2$) with a log
correction in relation (\ref{eq:1}) \cite{2Ghosh2022}).

An important observation (see Figs. \ref{fig:1}) has been the Gamma function
for the distribution  of the $h$ indices for all these (arbitrarily)
divided six ranges of top scorers.   This  indicates a
Chakraborti-Chakrabarti  type  kinetic exchange model \cite{8Chakraborti,9Sen2014} of citation
dynamics for each new paper, with a random  citation sharing fraction over a fixed (saved)
faction of citations of the close-circle papers. This ``saving" fraction determines (see e.g., \cite{10Patriarca2004}) the exponent
$\gamma_h$ in the distribution (\ref{eq:3a}) and the conservation of the total
citations in such ``social dynamics" of citations is practically determined
by the total publications within the ``aging'' period (see e.g., \cite{11Basu2005} and the
references therein). Indeed, for such a Gamma distributed statistics (\ref{eq:3})
in the Chakrabort-Chakrabarti kinetic exchange model (with fixed fraction
close circle citation propensity), the  analysis of  Patriarca et al. 
\cite{10Patriarca2004} suggests the relation (\ref{eq:4}). Such a relation fits extremely well with the values of the noise level (temperature) $T$ obtained by fitting the Hirsch index distribution data to the relation (\ref{eq:4}) and the value of $h_{av}$ obtained from distribution (\ref{eq:3a}) of $h$ together with its $\gamma_h$ value. As mentioned already, this indicates an effective kinetic exchange like stochastic dynamics for citations where each author has a fixed share of core-group citations and allows the rest from the literature. The dynamics give the total citations per paper constant on an average (constant value weakly dependent on the c-score rank or the Group).

In fact, the relation (\ref{eq:5})
fits very well with the data set for each Group with $\alpha_p=0.5$ (see Figs. \ref{fig:5}). Combining relations  (\ref{eq:1}) and (\ref{eq:5}) with $\alpha_c=0.5=\alpha_p$ and $D_c=0.5$, one gets the average citations per paper $N_c/N_p$ or the average coordination number of the citations network equal to $4D_p^2$, which ranges from 58 to 29 (see Table \ref{tab:2}). This was observed and reported earlier \cite{Ghosh2021} and can be viewed as an effective Dunbar number \cite{Dunbar1992} for the citations network.

Unlike the fitting value (0.50; see Table \ref{tab:2} ) of the prefactor $D_c$
in eqn (\ref{eq:1}). The fitting values of the prefactor $D_p(r)$ in eqn. (\ref{eq:5})
increase with  the rank $r$ (see table  \ref{tab:2}).  Fig. \ref{fig:6} gives the
extrapolated value of $D_p$ for the top rank ($r = 1$) to be  about
4.31,   which gives the  limiting value of the  citation network
coordination number (network average of citations per paper) to be
$4[D_p (r=1)]^2 \simeq 75$.

\section{Summary and Conclusion}
We analyze the distributions $f(h)$, $f(N_c )$ and
$f(N_p)$ of the Hirsch index ($h$), total citations
($N_c$) and total number of papers ($N_p$) of the
top 120,000 scorers (scientists with c-score in the
range  5.6125 to 3.3461) from the Stanford cite-score 2022
list, dividing them into six successive and equal sized
Groups,  and their corresponding $h$  ($ 3 \le h \le
284$), $N_c$  ($1009\le N_c \le 428620)$ and $N_p$ ($3\le N_p \le 3791$) from the Scopus
data. It may be mentioned that all these authors fall within (indeed the  toppers of) the top 2\% scientists in the Stanford cite-score (2022) selection list
\cite{6scopus,7elsevier}. As may be seen from Table \ref{tab:2}, while
fitting the $h$ index data to those for $N_c$ or $N_p$, using
the power law relations (\ref{eq:1}) and (\ref{eq:5}) respectively,
we found that while the powers $\alpha_c$ and
$\alpha_p$ both assumes the same value (1/2), and
the value of $D_c$ remains also 0.5 across  the
Groups, the value of $D_p$ varies very slowly with
the c-score value of the scientist. In order to
get sufficient statistics, yet to capture the changes in the values of $D_p$, we divided (arbitrarily)
the scientists into six equal Groups  (I, II, III
IV, V and VI), each having 20,000 scientists
according to their successive c-score ranks.
We find in each Group $f(h)$, $f(N_c)$ and $f(N_p)$ fit well with Gamma function form (\ref{eq:3a}), (\ref{eq:3b}) and (\ref{eq:3c}) (see Figs \ref{fig:1}, \ref{fig:2}, and \ref{fig:3}), e.g., $f(h) \sim h^{\gamma_h}[\mathrm{exp}(-h/T_h)]$, with the exponent
$\gamma_h \simeq 11.0 $, $\gamma_c \simeq 3.0 $ and $\gamma_p \simeq 2.2 $  and the noise levels $T_h$, $T_c$ and $T_p$ dependent on the c-score range
considered.  
We compared the data (directly obtained from
Scopus) for $f(h)$ in all the six Groups (see
Figs. \ref{fig:1}) with those obtained from the data
for $f(N_c)$, using the relation (\ref{eq:1}) with $\alpha_c$
= 1/2 and $D_c = 0.5$, and got excellent overlap (see
Figs. \ref{fig:4}). We did also same, obtained from the
$f(N_p)$ data, using the relation (\ref{eq:5}), again with
$\alpha_p =1/2$, but $D_p$ dependent on the Group
(see Table \ref{tab:2}), showing again very good fit (see
Figs. \ref{fig:5}).
Other suggestions like $\alpha_c \simeq 0.42$ \cite{5Radicchi2013}
or $\alpha_c = 0.5$  but with an inverse $\mathrm{log}N_c$
correction term \cite{2Ghosh2022} do not give good fits (see the
insets of Fig. \ref{fig:4}). In fact, a very recent extensive
analysis \cite{biro2023} of the statistical relation (\ref{eq:1})
between $h$ and $N_c$ from the Google Scholar data
gave $\alpha_c = 0.5$ and $D_c$ = 0.5 (as obtained
here), confirming a much earlier study \cite{Glanzel2006}. In our
study, apart from similar relationship between
$h$ and $N_p$ and the relationships (\ref{eq:3a}-\ref{eq:3c}), we
find here the $T_h$ values (in relation (\ref{eq:3a})) for
each of the six c-score ranges fit very well with the
relation $T_h = h_{av}/(\gamma_c + 1)$ where  $h_{av}$ is the average of $f(h)$ in each Group. This compares very well with the  Chakraborti-Chakrabarti model \cite{8Chakraborti,10Patriarca2004,Pareschi2014} of ``wealth"  distribution where a fixed saving
fraction of the wealth (which determines the value of the exponent $\gamma_c$  in the Gamma distribution) is retained in each kinetic
exchange or interaction,  suggesting a similar stochastic
dynamics of paper citations, where the fixed fraction of
(confident or core Group) ``citations" (wealth) in each paper-writing (interaction) determines the exponent $\gamma_c$ value and the
corresponding
noise level $T_h$ in $f(h)$. We also observe an interesting feature of
the citation
network. The observation (relations (\ref{eq:1}) and (\ref{eq:5})) $h =D_c N_c^{\alpha_c}= D_p N_p^{\alpha_p}$, where  $\alpha_c=\alpha_p$ =
0.5, $D_c=0.5$ and $2.7
\le D_p \le 3.8$ depending on the Group, suggesting  the value ($N_c/N_p=(D_p/D_c)^2=4D_p^2$) of the
average citation
per paper shown in Table \ref{tab:2} 
depends on the Group the scientist belongs to and
ranges from 58 (for Group I) to 29 (for Group IV). As
discussed at the end of the last section (see Figs. \ref{fig:6}),
the limiting value of this citation-network coordination
number (network average of citations per paper) gets
extrapolated to about 75.  This then may be identified
as the effective social coordination number (or  Dunbar
number \cite{Dunbar1992})  giving the group size of the top-rated
scientists today, as in the primates.

\section*{Acknowledgement}
We are thankful to Soumyajyoti Biswas and Parongama Sen for several useful comments on the manuscript. BKC is grateful to the Indian National Science Academy for their Senior Scientist Research Grant.

\end{document}